\documentclass[%
reprint,
 amsmath,amssymb,
 aps,
pra,
]{revtex4-1}

\usepackage{graphicx}% Include figure files
\usepackage{dcolumn}% Align table columns on decimal point
\usepackage{bm}% bold math
\usepackage{float}
\usepackage{hyperref}
\begin{document}

%\preprint{APS/123-QED}

\title{Coupled quantum dots in bilayer graphene}

\author{Marius Eich}
 \email{Electronic mail: meich@phys.ethz.ch}
\author{Riccardo Pisoni}
\author{Alessia Pally}
\author{Hiske Overweg}
\author{Annika Kurzmann}
\author{Yongjin Lee}
\author{Peter Rickhaus}
\author{Klaus Ensslin}
\author{Thomas Ihn}

\affiliation{%
 Solid State Physics Laboratory, ETH Zurich, 8093 Zurich, Switzerland \\
}

\author{Kenji Watanabe}
\author{Takashi Taniguchi}
\affiliation{%
 Advanced Materials Laboratory, NIMS, 1-1 Namiki, Tsukuba 305-0044, Japan \\
}

\date{\today}% It is always \today, today,

\begin{abstract}
Electrostatic confinement of charge carriers in bilayer graphene provides a unique platform for carbon-based spin, charge or exchange qubits. 
By exploiting the possibility to induce a band gap with electrostatic gating, we form a versatile and widely tunable multi-quantum dot system.
We demonstrate the formation of single, double and triple quantum dots that are free of any sign of disorder. 
In bilayer graphene we have the possibility to form tunnel barriers using different mechanisms. 
We can exploit the ambipolar nature of bilayer graphene where pn-junctions form natural tunnel barriers.
Alternatively, we can use gates to form tunnel barriers, where we can vary the tunnel coupling by more than two orders of magnitude tuning between a deeply Coulomb blockaded system and a Fabry-P\'{e}rot-like cavity.
Demonstrating such tunability is an important step towards graphene-based quantum computation.
\end{abstract}

\maketitle

\section{\label{sec:Intro}Introduction}

Soon after its discovery, graphene was recognized as a promising platform for spin qubits because of carbon's low atomic mass leading to small spin-orbit coupling and the small hyperfine interaction in carbon-based devices \cite{trauzettel_spin_2007}.
Since the monolayer system does not exhibit a band gap and therefore does not allow for electrostatic confinement of charge carriers, early experiments relied on defining graphene nanostructures by etching monolayer graphene \cite{ponomarenko_chaotic_2008,guttinger_transport_2012}.
These devices suffered from charge carrier localization at the rough sample edges \cite{bischoff_localized_2015}.
In contrast, bilayer graphene offers the possibility to open a band gap by applying a displacement field normal to the bilayer plane \cite{mccann_asymmetry_2006,ohta_controlling_2006,oostinga_gate-induced_2008}.
With a suitable design of top- and back-gate electrodes, this allows for electrostatic confinement of charge carriers in high quality bilayer graphene devices \cite{goossens_gate-defined_2012,allen_gate-defined_2012}.
Until recently, devices utilizing electrostatic confinement of charge carriers suffered from low pinch-off resistances that were achieved when opening a band gap and tuning the Fermi energy into the gap \cite{goossens_gate-defined_2012,allen_gate-defined_2012,zhu_edge_2017}.
The implementation of a graphite back-gate significantly improved the pinch-off resistances in electrostatically defined bilayer graphene nanostructures reaching values many orders of magnitude above the resistance quantum \cite{zibrov_tunable_2017,overweg_electrostatically_2018}.
We suspect that the graphite back-gate enhances the homogeneity of the charge carrier density in van der Waals heterostructures by screening the influence of impurities and defects in the oxide layer of the silicon substrates on the electron gas and thus reducing the amplitude of the disorder potential.

Forming quantum-dot-based qubit systems relies on controlling the occupation of the quantum dots (QDs) on the few-electron level and in the single-level transport regime. 
The device structure developed in Ref.~\citenum{overweg_electrostatically_2018} allowed us to confine both single electrons and holes electrostatically in bilayer graphene \cite{eich_spin_2018}.
State of the art qubit devices in GaAs or silicon rely on coupling multiple QDs in series to create charge \cite{gorman_charge-qubit_2005,viennot_out--equilibrium_2014}, spin \cite{petta_coherent_2005,cottet_spin_2010}, spin-orbit \cite{tokura_coherent_2006,nadj-perge_spinorbit_2010}, exchange \cite{gaudreau_coherent_2012,medford_quantum-dot-based_2013,russ_coupling_2016}, hybrid \cite{shi_fast_2012,thorgrimsson_extending_2017} or valley-orbit \cite{mi_electrically_2018} qubits.
Here we show that the design developed in Refs.~\citenum{overweg_electrostatically_2018} and \citenum{eich_spin_2018} can be used to create widely tunable multi-dot systems in bilayer graphene.
The complexity of the multi-dot system is determined by the number of top-gates that are fabricated on top of the van der Waals heterostructure, such that the design can be scaled up to create fully tunable multi-dot systems.
Bilayer graphene QD systems combine the flexibility of a two-dimensional platform similar to two-dimensional electron gases in GaAs heterostructures \cite{kouwenhoven_few-electron_2001} with the option to couple QDs of opposite charge carrier polarity similar to carbon nanotube systems \cite{cobden_shell_2002,jarillo-herrero_electron-hole_2004,pei_valleyspin_2012}.

In this paper we start by characterizing the device and introducing the various coupled QD regimes that can be realized.
Afterwards we will use six gates (graphite back-gate, two split gates and three finger gates) to define a fully tunable single-QD, where the tunnel coupling of the QD to the leads is controlled electrostatically - an important ingredient for defining multi-QD qubit systems \cite{gorman_charge-qubit_2005,petta_coherent_2005,tokura_coherent_2006,cottet_spin_2010,nadj-perge_spinorbit_2010,shi_fast_2012,gaudreau_coherent_2012,medford_quantum-dot-based_2013,viennot_out--equilibrium_2014,russ_coupling_2016,thorgrimsson_extending_2017,mi_electrically_2018}.
The plunger gate of the single-QD can be used to deplete the central region of the QD, leading to a crossover into a double-QD regime, where the individual QDs have the same charge carrier polarity.
Utilizing the ambipolar nature of graphene and the natural tunnel barriers that arise at pn-junctions \cite{eich_spin_2018} we can tune the device into a pnp-triple-dot regime and more complex multi-QD regimes.

\section{\label{sec:Char}Characterization}

We investigated a bilayer graphene device featuring a graphite back-gate, the bilayer graphene flake encapsulated between two boron nitride flakes and two layers of top-gate electrodes separated by an insulating layer of aluminum oxide. 
The van der Waals heterostructure was assembled and contacted as in Refs.~\citenum{overweg_electrostatically_2018} and \citenum{eich_spin_2018}.
The inset in Fig.~1(a) shows the sample and in particular the top-gate structure (for fabrication details see Ref.~\citenum{eich_spin_2018}).
A pair of split gates (brown) spans the bilayer graphene flake (red dashed outline).
By applying a positive voltage to the graphite back-gate and a negative voltage to the pair of split gates, a strong displacement field is applied to the bilayer graphene regions below the split gates.
The strong displacement field opens a band gap in these regions \cite{mccann_asymmetry_2006,ohta_controlling_2006,oostinga_gate-induced_2008} and with the Fermi energy being tuned into the gap, charge carriers are laterally confined and flow through the $\sim 160 \ \mathrm{nm}$ wide channel between the split gates.

Three finger gates (L, M and R) on top of the split gates, separated by a layer of aluminum oxide, cross the channel and are used to control the charge carrier density inside the channel locally \cite{overweg_electrostatically_2018,eich_spin_2018}. 
The finger gates are $\sim 40 \ \mathrm{nm}$ wide and separated by $\sim 120 \ \mathrm{nm}$.
The outer grayed-out finger gates seen in the inset of Fig.~1(a) were not used during the experiment, but can be utilized to tune the device into more complex coupled QD regimes than those described in this paper. 
All the presented measurements were performed in a dilution refrigerator with a base temperature of $10 \ \mathrm{mK}$ in a two-terminal DC setup with a symmetrically applied bias voltage.

To characterize the device, we measure the conductance at $V_{SD} = 100 \ \mathrm{\mu V}$ of bias (note that $V_{SD} \gg k_{B}T$ to increase the signal-to-noise ratio) and $V_{M} = 0 \ \mathrm{V}$ as a function of the voltages $V_{L}$ and $V_{R}$ applied to gates L and R, respectively, shown in Fig.~1(a).
The map shows three distinct sets of resonances and can be divided into four quadrants (\textrm{I to IV}).
Zooms into the four quadrants are shown in Figs.1(b-e) with the corresponding sketch of the charge density distribution along the channel shown above the conductance maps. 
Quadrant \textrm{I} corresponds to a gate-defined single n-type QD. 
Gate L (R) is tuned close to charge neutrality, thus creating a tunnel barrier between source (drain) and the n-type QD forming between gates L and R. 
In this regime, discussed in more detail later, gate M can be used as the plunger gate of the QD.
The capacitance between gates L, R and the middle QD leads to resonances with a slope of $-1$. 
The respective lever arms are $\alpha_{L}^{M} = 0.17$ and $\alpha_{R}^{M} = 0.16$. 

As soon as the voltage on gate L (R) is negative enough to form a p-type QD below itself \cite{eich_spin_2018}, a transition from the single-dot regime \textrm{I} to the double-dot regime \textrm{II} (\textrm{IV}) is observed. 
The zooms in Figs.~1(b,e) show a double-dot charge stability diagram with the expected hexagonal pattern indicated by dotted lines.
We observe that the charging lines of the QD forming between gates L and R are more prominent, which means that transport through the system is dominated by cotunneling processes via the outer p-type QDs.

\begin{center}
\begin{figure}[t]
\includegraphics[width=0.45\textwidth]{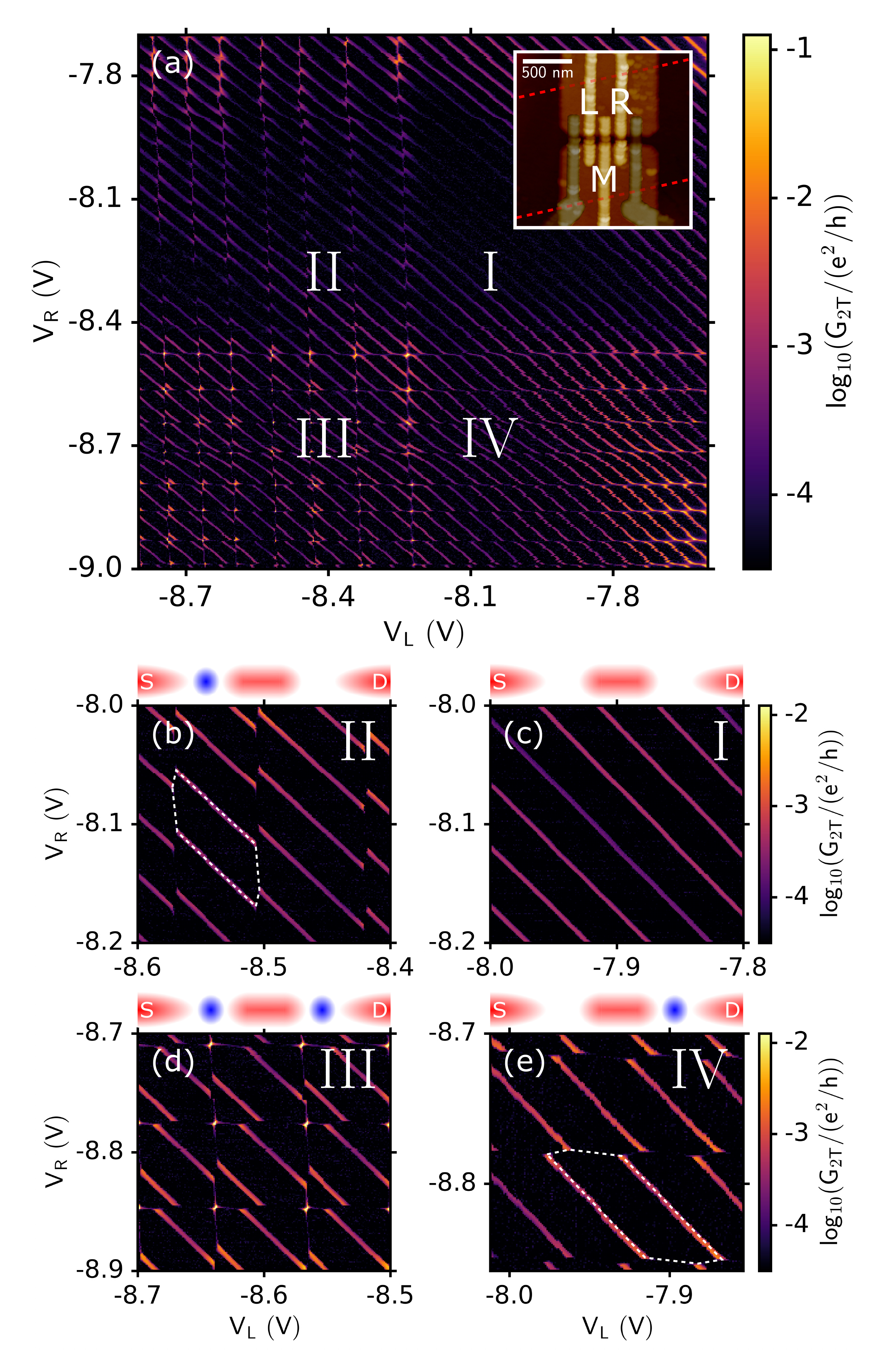}
\caption{
Characterization of the device.
(a)~Logarithmic two-terminal conductance as a function of the left and right finger gate (middle finger gate is unbiased). 
The conductance map can be divided into four regions: The single-dot regime (\textrm{I}) with a zoom-in show in (c), the two pn-double-dot regimes (\textrm{II}) and (\textrm{IV}) with zoom-ins shown in (b) and (e), respectively, and the pnp-triple-dot regime (\textrm{III}) with a zoom-in shown in (d). 
Schematics on top of (b-d) indicate n-type (red) and p-type (blue) QDs connected to n-type source (S) and drain (D).
Dashed lines in (b,e) outline the hexagonal structure of the charge stability diagram in the double-dot regimes.
Inset: Scanning force micrograph of the split gate (brown) forming the channel and finger gate (yellow) structure on top of the encapsulated bilayer graphene flake (dashed lines).
Grayed-out gates are not used, while the voltages applied to gates labeled L, M, and R are varied.  
}
\end{figure}
\end{center}
\vspace{-6ex}

\indent When a p-type QD forms below both gates L and R, we reach the pnp-triple-dot regime in quadrant \textrm{III}. 
The zoom-in shown in Fig.~1(d) shows three distinct sets of resonances: vertical (horizontal) resonances of the p-type QD below gate L (R) and diagonal resonances of the n-type QD in the middle. 

The schematic in Fig.~2(a) shows the situation for a triple-dot formed along the channel.
The positive back-gate voltage ($V_{BG} = 5 \ \mathrm{V}$) induces a finite electron density inside the channel.
By applying a negative voltage to any of the finger gates, the electron density underneath the respective gate can be reduced.
At some point, the electron gas below the finger gate is completely depleted (Fig.2(b)) and transport through the channel is pinched off.
Reducing the finger gate voltage further, a finite hole density is induced, forming small p-type island inside the n-type channel (Fig.2(c)). 
The resulting pn-junctions provide natural tunnel barriers such that a p-type QD can be formed under each of the finger gates \cite{eich_spin_2018}.
In quadrant \textrm{III} of Fig.~1(a) two p-type QDs are formed below gates L and R which are tunnel-coupled via pn-junctions to source and drain as well as the larger n-type QD forming between the gates. 
The respective band alignment at different positions along the channel is shown in the bottom part of Fig.~2(a). 
For gates L and R, the applied voltage is sufficiently negative to pull the top of the valence band above the Fermi energy, creating a p-type dot underneath the gates. 
For source and drain, as well as the region between gates L and R, the Fermi energy lies in the conduction band. 
At the transition between n-type and p-type regions, the Fermi energy has to lie in the band gap creating a finite region of zero charge carrier density \cite{eich_spin_2018}.
The details of the triple-dot regime will be discussed later in this manuscript.

In addition to the pnp-triple-dot, Figs.~2(b-n) schematically show the other QD systems that can be realized with three finger gates on top of the channel.
The situation shown in Fig.~2(b) corresponds to a quantum point contact which is discussed in Ref.~\citenum{overweg_electrostatically_2018}.
Forming a single-gate-defined QD as shown in Fig.~2(c) is discussed in Ref.~\citenum{eich_spin_2018}.
The four quadrants in Fig.~1(a) correspond to the configurations of Figs.~2(a,d,f).
Later in the manuscript we will present data on the double-QD regime shown in Fig.~2(h), while additional data on the remaining configurations of Fig.~2 can be found in the supporting information.
It becomes apparent that the presented device structure allows for tuning the system into a highly complex multi-dot system, where $N$ finger gates can define a serial system of $2N-1$ QDs of opposite polarity.
In addition, the voltages applied to the graphite back-gate and the split gates can be inverted with respect to the overall Dirac point of the device, forming a channel with a finite hole density.
Each of the configurations shown in Fig.~2 can therefore also be realized with inverted charge carrier polarity.
In contrast to traditional semiconductor heterostructures, gate-defined bilayer graphene nanostructures facilitate the investigation of coupled multi-dot systems with varying charge carrier polarity.
Similar systems can be realized in carbon nanotubes \cite{pei_valleyspin_2012}, but bilayer graphene offers the flexibility of a two-dimensional platform.

\begin{center}
\begin{figure}[t]
\includegraphics[width=0.45\textwidth]{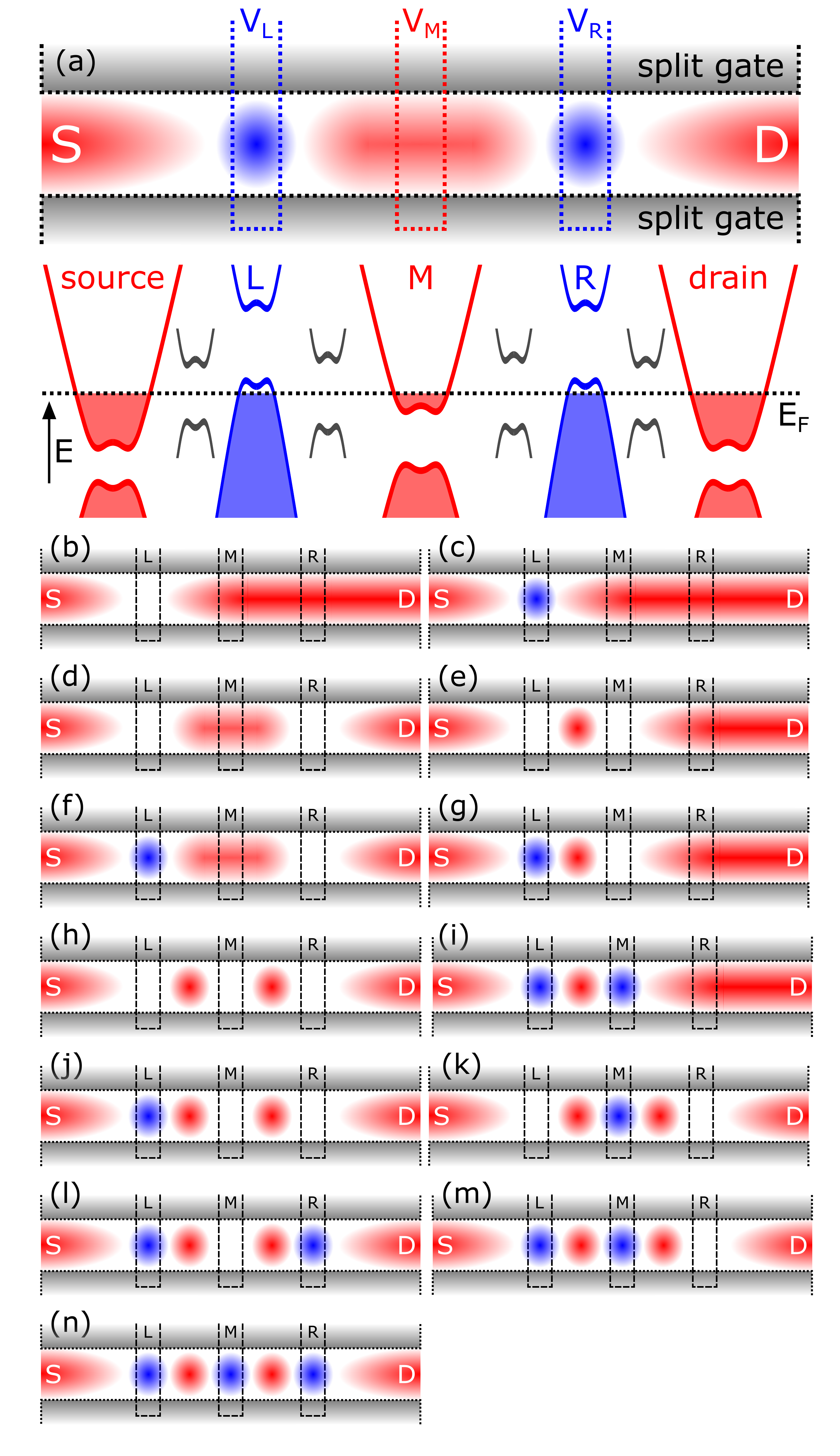}
\caption{
Possible device configurations.
(a)~Top: Schematic of the split and finger gates. 
Red (blue) color represent a finite electron (hole) density inside the channel, here shown for the triple-dot regime. 
Bottom: Schematic representation of the band structure at different positions along the channel. 
(b-n)~Possible additional configurations of the device for varying charge carrier polarity under gates L, M and R for an n-type channel.
Note that the polarity can be inverted by inverting all applied gate voltages with respect to the overall Dirac point of the device.
}
\end{figure}
\end{center}

%\vspace{4ex}
\section{\label{sec:SingleQD}Fully tunable single-dot}

\begin{center}
\begin{figure}[htb]
\includegraphics[width=0.45\textwidth]{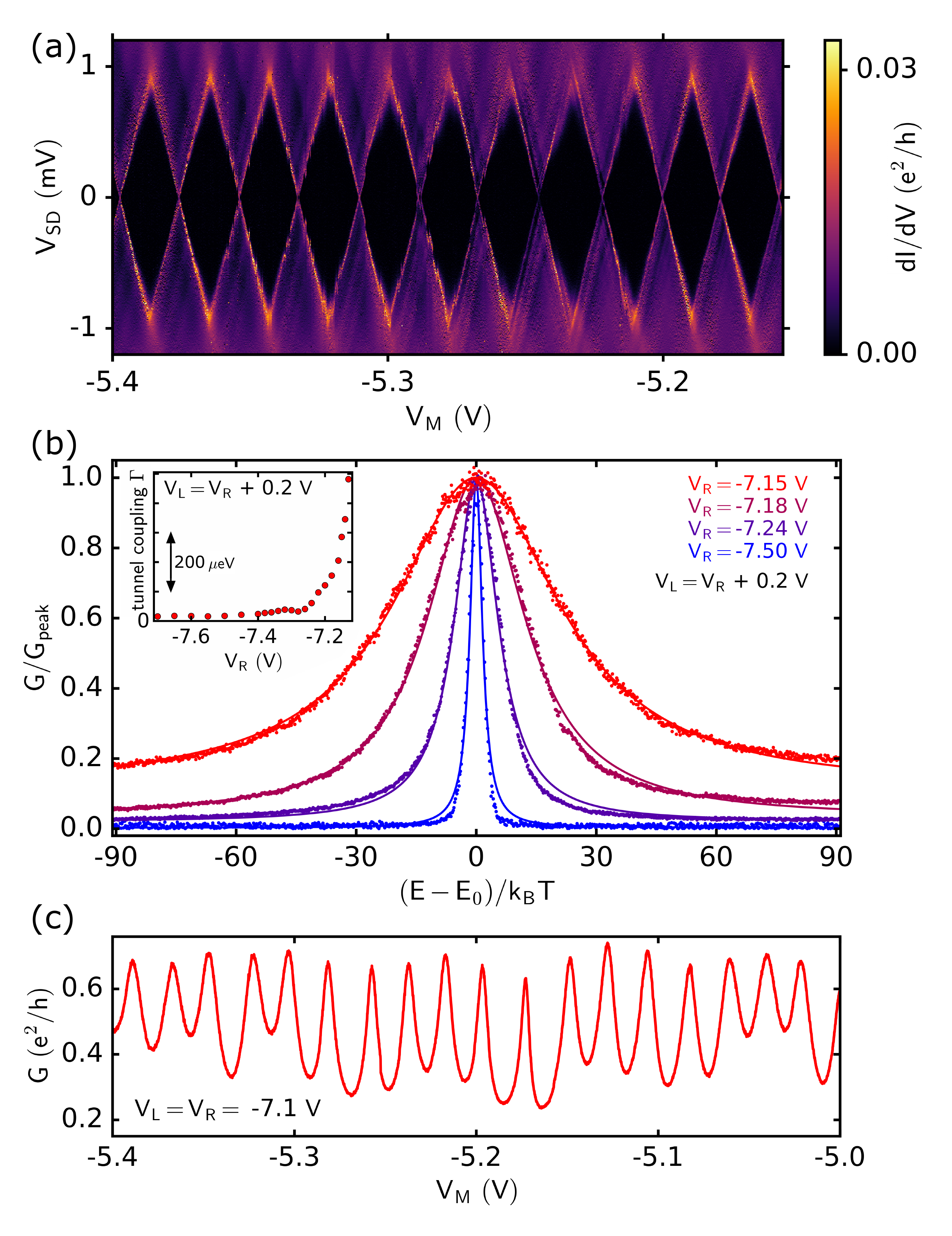}
\caption{
Gate-defined electron QD.
(a)~Coulomb diamonds in the single-dot regime with gates L and R defining the tunneling barriers ($V_{L} = -7.6 \ \mathrm{V}$ and $V_{R} = -7.55 \ \mathrm{V}$) and gate M acting as the plunger gate.
%The extracted charging energy is $920 \ \mathrm{\mu eV}$. 
(b)~Coulomb resonances for varying tunnel coupling tuned by gates L and R. 
The plunger gate axis is converted to energy using the lever arm of gate M extracted from (a) and $T = 50 \ \mathrm{mK}$.
Solid lines represent Lorentzian fits to the data. 
Inset: Tunnel coupling extracted from the fits as a function of the voltage applied to gate L. 
Note that $V_{L} = V_{R} + 0.2 \ \mathrm{V}$.
(c)~Conductance trace in the case of an almost completely open system similar to a Fabry-P\'{e}rot interferometer.
}
\end{figure}
\end{center}

\vspace{-5ex}
The n-type single-QD seen in quadrant \textrm{I} of Fig.~1(a) and shown schematically in Fig.~2(d), can be understood in close analogy to conventional GaAs QDs \cite{fujisawa_inelastic_2000}. 
With the channel being defined by the split gates, gates L and R define the tunnel barriers coupling the QD to the leads, while gate M serves as the plunger gate.
Gates L and R are tuned close to their respective charge neutrality point such that the regions underneath the gates are depleted forming tunnel barriers to the n-type QD in the middle.
In contrast to the single-gate-defined QDs presented in Ref.~\citenum{eich_spin_2018} where the tunnel coupling is an intrinsic property of the pn-junctions, defining the QD with three gates allows us to tune the tunnel coupling.
By depleting the electron density underneath gates L and R and forming tunnel barriers to the QD in-between, the resistance of the system exceeds $10 \ \mathrm{G\Omega}$ (when the QD is off resonance), which proves the high quality of the electrostatically induced barriers.
The plunger gate in turn allows us to change the occupation number of the QD while keeping the tunnel coupling constant, thus granting full control over the QD \cite{ihn_semiconductor_2009}.
By raising (lowering) the voltage $V_{M}$, individual electrons can be added to (removed from) the QD, leading to a very regular sequence of Coulomb resonances (see Fig.~3(a) and Fig.~S1).
Since this QD is occupied by a large number of electrons, the addition energy stays constant over a large range of $V_{M}$ \cite{kouwenhoven_few-electron_2001}.
The charging energy of the QD is $E_{ch}^{M} = 920 \ \mathrm{\mu eV}$ which can be extracted from the Coulomb diamond measurement shown in Fig.~3(a) and the lever arm of the plunger gate M is $\alpha_{M}^{M} = 0.042$.
To reach the few-charge carrier regime in future devices, the plunger gate should cover the whole area of the QD, for example by adding an additional layer of Al$_2$O$_3$ and overlapping the plunger gate with the finger gates defining the tunnel barriers.

In Fig.~3(b) we plot Coulomb resonance peaks for different tunnel coupling $\Gamma$ to the leads. 
The solid lines represent Lorentzian fits \cite{beenakker_theory_1991} to the data.
For high tunnel coupling the fits perfectly match the data while at low tunnel coupling the slight deviation is due to the thermal broadening of the peaks, which becomes increasingly significant with deceasing tunnel coupling. 
For individual Coulomb resonances the extracted tunnel coupling varies by less than 10\%.
The inset shows the extracted average tunnel coupling for varying voltage applied to gates L and R tuning the tunnel barriers, demonstrating that we can change the tunnel coupling over two orders of magnitude.
For the lowest tunnel coupling we can tune the QD into a regime where the broadening of Coulomb resonances is limited by the electronic temperature.
The extracted upper bound to the electronic temperature is $T = 50 \ \mathrm{mK}$.
Measuring Coulomb resonances as a function of increasing temperature shows a decreasing peak value of the conductance resonance, which indicates that transport through the QD occurs via a single-QD energy level (see supporting information).
In contrast, for very high tunnel coupling of the QD to the leads, we measure the conductance trace shown in Fig.~3(c). In this regime, the system can also be viewed as a Fabry-P\'{e}rot interferometer\cite{grushina_ballistic_2013,rickhaus_ballistic_2013,varlet_fabry-perot_2014}.

\section{\label{sec:DoubleDot}n-n double-dot}

\begin{center}
\begin{figure}[htb]
\includegraphics[width=0.45\textwidth]{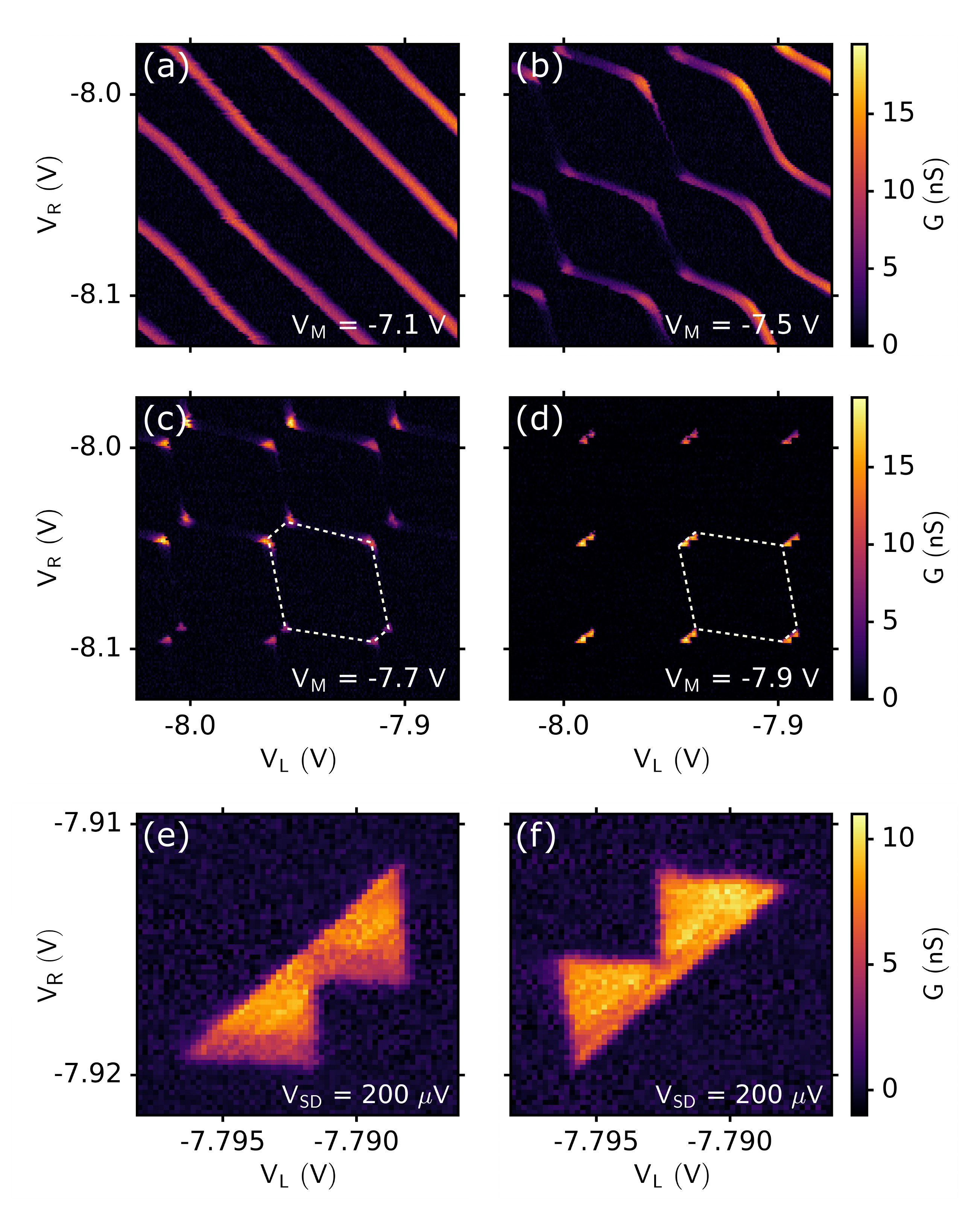}
\caption{
nn-double-QD.
(a-d)~Conductance map as function of the voltages applied to gates L and R. 
As the voltage applied to gate M is lowered from (a) to (d), the single-QD is split in two and the tunnel coupling between the two resulting electron QDs is continuously lowered. 
Dashed lines in (c,d) outline the hexagonal structure of the charge stability diagram of the resulting nn double-dot.
(e) Conductance at one exemplary triple point for positive and (f) negative applied bias with $V_{M} = -8 \ \mathrm{V}$.
}
\end{figure}
\end{center}
\vspace{-5ex}

Starting with the single n-type QD described above, we now take a closer look at the influence of gate M. 
For voltages above $-7 \ \mathrm{V}$ applied to gate M, states of the n-type QD are extended over the whole area between gates L and R.
However, by lowering the voltage $V_{M}$, the weight of the wave function underneath gate M will decrease and for sufficiently negative voltages $V_{M}$, the n-type dot will be split in two smaller serial QDs (schematically shown in Fig.~2(h)) separated by a single tunnel barrier.
This evolution from a single-dot to a tunnel-coupled double-dot \cite{livermore_coulomb_1996} is shown in Fig.~4(a-d) where the voltage applied to gate M is continuously lowered.
In Fig.~4(a) we see the same situation of a single n-type QD as in Fig.~1(c).
The Coulomb resonances have a slope of $-1$ since the energy levels of the QD are equally tuned by gates L and R.

When tuning gate M closer to charge neutrality by lowering $V_{M}$, the characteristic hexagonal charge stability diagram pattern of the double-dot gradually develops as shown in Fig.~3(b). 
The two QDs form between gates L and M (dot LM), and gates M and R (dot MR).
Coulomb diamonds of these two individual QDs are shown in the supporting information, where we extract charging energies $E_{ch}^{LM} = 2.15 \ \mathrm{meV}$ and $E_{ch}^{MR} = 2.25 \ \mathrm{meV}$, respectively.
The increased charging energy compared to the larger single-QD spanning the whole area between gates L and R agrees well with the increased confinement for the two emerging QDs.
Due to the strong tunnel coupling between the two QDs forming the serial double-dot, the charging lines of the individual two QDs are clearly visible in Fig.~4(b).
As the tunnel coupling between the dots is decreased by lowering $V_{M}$ further, the charging lines gradually disappear (see Fig.~4(c)) until we reach the regime, where transport through the double-dot is only significant at the triple points between three stable charge configurations forming a transport cycle \cite{waugh_single-electron_1995,livermore_coulomb_1996}, shown in Fig.~4(d). 

The triple points are arranged on almost vertical and horizontal lines indicating small cross-capacitance between the left (right) QD and gate R (L).
Zooming into one of the triple points, we resolve finite bias triangles for positive (Fig.~4(e)) and negative applied bias (Fig.~4(f)). 
As expected, the boundaries of the triangles are given by almost vertical and horizontal lines as well as the diagonal interdot charging line which has a slope of one.
Transport though the double-dot within the triangles is dominated by inelastic tunneling.
For the present device we see no evidence for spin blockade \cite{ono_current_2002,pei_valleyspin_2012} over all investigated parameter regimes.
We anticipate that spin blockade will be observed in future double-QD devices where the constituent single-QDs can be tuned into the few-charge carrier regime.

\section{\label{sec:TripleDot}pnp-triple-dot}

\begin{center}
\begin{figure}[b]
\includegraphics[width=0.45\textwidth]{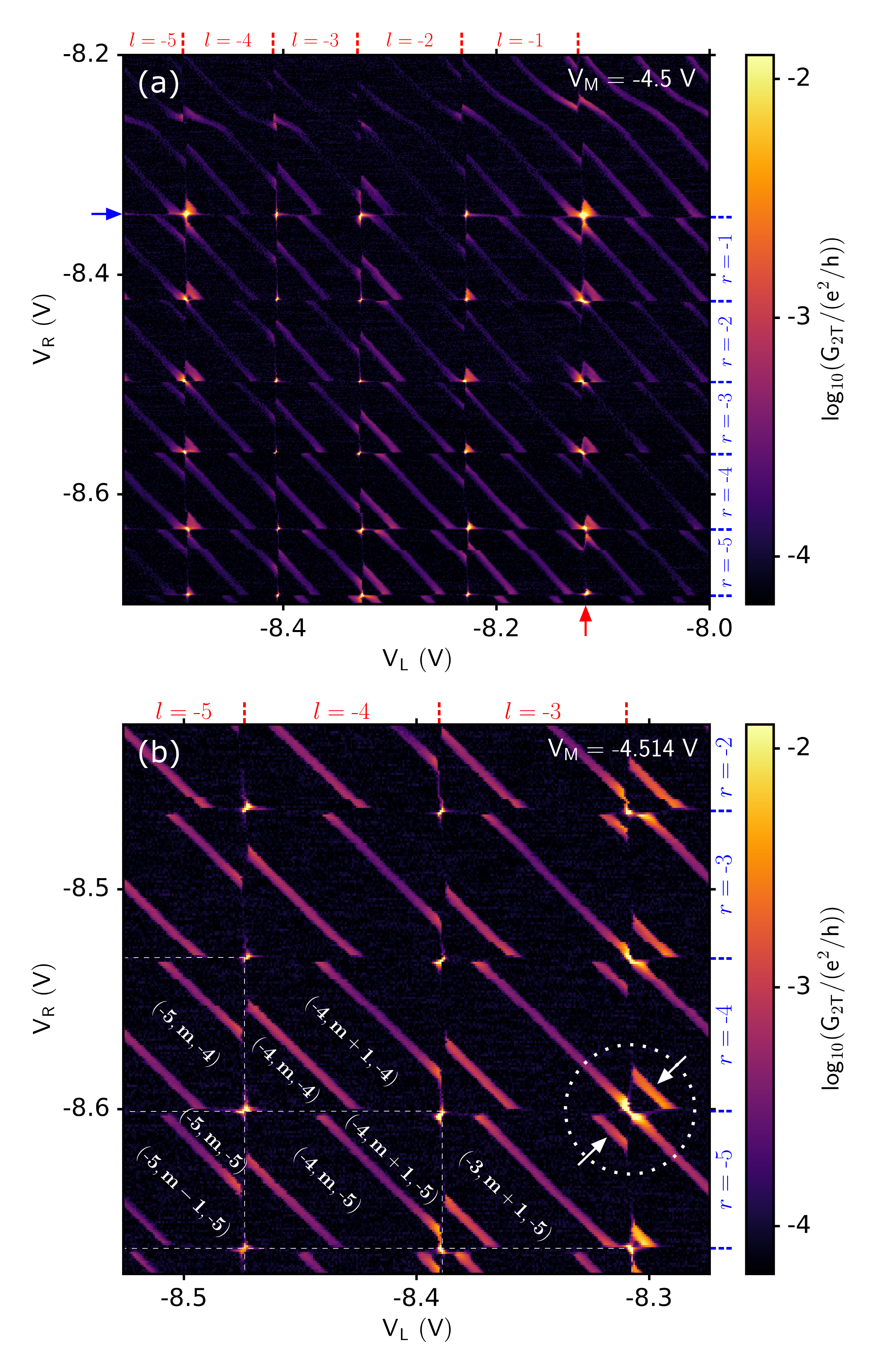}
\caption{
pnp-triple-QD.
(a,b)~Charge stability diagram of the pnp-triple-QD. 
Numbers on the top (right) side of the conductance maps indicate the hole occupation of the left (right) p-type QD under gate L (R).
The red (blue) arrow in (a) indicates charging of the first hole.
Triplets ($l$,$m$,$r$) in (b) indicate the triple-dot charge state at different positions in the charge stability diagram. Charging lines of the outer two dots are highlighted by dashed lines for better visibility.
The dotted white circle marks a crossing point where the levels of the three QDs are almost perfectly aligned.
White arrows indicate two replicas of the diagonal resonance of the middle QD.
}	
\end{figure}
\end{center}

\vspace{-5ex}
The triple-dot regime of the third quadrant in Fig.~1(a) is reached when a p-type QD forms below both gate L and R while gate M is above -7 V, where it does not split the n-type single-dot into a double-dot. 
A closer overview of this regime, which is schematically shown in Fig.~2(a), is presented in Fig.~5(a).
The point where the first hole is charged onto the p-type QD under gate L (R) is indicated by a red (blue) arrow.
For decreasing voltage $V_{L}$ ($V_{R}$) the respective dots are charged successively with individual holes.
The occupation number $l$ ($r$) of the two p-type QDs is indicated on the top (right), where negative integers refer to the occupation of the QD with holes.
In addition to the vertical and horizontal resonances stemming from charging of the outer p-type QDs, we observe diagonal resonances of slope $-1$ corresponding to charging of the middle n-type QD with single electrons. 
By changing the voltage applied to gate M, these resonances can be shifted with respect to the charging lines of the outer p-type QDs (see supporting information).

Figure~5(b) shows a zoom into the triple-dot charge stability diagram in the vicinity of $l = r = -4$.
White dashed lines in the bottom left corner of Fig.~5(b) highlight the resonances of the outer p-type QDs.
For the outer p-type QDs, we are in the few-hole regime \cite{eich_spin_2018} which allows us to label each charge state with the respective occupation number of the outer QDs.
The middle n-type QD is in the many electron regime and we label its (unknown) occupation with $m$.
By increasing $V_{L}$ and $V_{R}$ simultaneously and crossing a diagonal resonance of the middle QD, its occupation number is increased by $1$.

In the situation where the energy levels of all three QDs are alomst perfectly aligned (energy level crossing highlighted by dotted circle in Fig.~5(b)) we see two replicas (indicated by white arrows) of the diagonal resonance of the middle dot. 
When crossing the point of threefold level alignment along a diagonal line with slope one, both outer p-type QDs are charged with an additional electron.
The capacitive coupling of the outer QDs and the middle QD shifts the energy levels of the middle QD up such that the occupation number of the middle QD decreases by $1$. 
Overall, the system is therefore again charged with one additional electron \cite{schroer_electrostatically_2007}.
The middle QD is then charged again with an electron at higher $V_{L,R}$.
A detailed discussion of the charge stability diagram close to the threefold level crossings and the influence of gate M is included in the supporting information.

\section{\label{sec:Conc}Conclusion}

The presented results prove the versatility of charge carrier confinement in electrostatically defined bilayer graphene nanostructures.
With charge carriers being confined to a narrow channel, three finger gates locally tune the charge carrier density inside the channel leading to a wide variety of multi-dot systems. 
With the presented single-QD results, we reach the same tunability as QD systems in GaAs or silicon. 
Further reducing the size of the system by designing neighboring finger gates with smaller spacing will allow for tuning these single-QDs into the few-electron or few-hole regime.
The charge stability diagrams of the double- and triple-dot systems studied in this paper is of comparable quality as the results obtained with coupled QDs in GaAs or silicon.

Traditional multi-dot systems in GaAs offer the flexibility of a two-dimensional platform, whereas carbon nanotube systems allow for coupling QDs of opposite polarity and exploiting the valley degree of freedom.
Both these advantages can be combined in our bilayer graphene devices and scalability can be achieved by using bilayer graphene grown by chemical vapor deposition \cite{schmitz_high_2017}.
Making use of the ambipolarity of the system, very complex structures like charge density checkerboard patterns may be realized with a limited amount of top-gates in the future.
With minimal alteration of the top-gate structure, the presented platform facilitates the study of valley-spin blockade \cite{pei_valleyspin_2012} in nn-, pn- and pp-double-QDs as well as QD arrays where the individual QDs could be manipulated by an additional top-gate layer.
The valley degree of freedom can be exploited to create qubits which are insensitive to charge noise \cite{mi_electrically_2018} but in contrast to silicon where the valley splitting results from strain during the growth, the valley splitting in our QDs can be tuned by a perpendicular magnetic field \cite{eich_spin_2018}.
In addition, the influence of the valley degree of freedom in bilayer graphene can be studied by designing systems with varying geometry aligned to the crystallographic orientation of the bilayer graphene flake \cite{gorbachev_detecting_2014}.
 
Electrostatically defined multi-dot systems in bilayer graphene open up a wide field of research where QDs of any polarity can be coupled in arbitrary sequence.
Fully tunable single- and double-QD systems will facilitate the search for Kondo and spin-blockade physics in bilayer graphene.
Single-, double- and triple-QD systems with full control over the tunnel coupling to the leads provide a promising platform for charge, spin, exchange, hybrid and valley-orbit qubits with potentially long coherence times.

While preparing the manuscript we became aware of another publication on double quantum dots realized in bilayer graphene \cite{banszerus_gate-defined_2018}.

\acknowledgments

We thank Peter M\"arki, Erwin Studer, as well as the FIRST staff for their technical support.
We also acknowledge financial support from the European Graphene Flagship, the Swiss National Science Foundation via NCCR Quantum Science and Technology, the EU Spin-Nano RTN network and ETH Zurich via the ETH fellowship program. Growth of hexagonal boron nitride crystals was supported by the Elemental Strategy Initiative conducted by the MEXT, Japan and JSPS KAKENHI Grant Number JP15K21722.

\noindent \textbf{Author contributions:} M.E. and R.P. fabricated the device. M.E. and A.P. performed the measurements. H.O., A.K., Y.L., and P.R. supported device fabrication and data analysis. K.W. and T.T. provided high-quality boron nitride crystals. K.E. and T.I. supervised the work. 
The authors declare that they have no competing interests.

\noindent \textbf{Supporting Information:} The Supporting Information is available free of charge on the \href{https://pubs.acs.org/doi/suppl/10.1021/acs.nanolett.8b01859}{Nano Letters webpage} and includes additional data on a single n-type dot, the constituent dots in the double-dot regime, triple-dot level crossings, the npn-triple-dot, a 2-gate triple-dot and the quadruple- and quintuple-dots.

\end{document}